\documentstyle[12pt]{article}
\begin{document}

\vspace{2cm}
\title{The Exact Solution of the $SU(3)$ Hubbard Model}
\author{Buoyu Hou$^1$ \thanks{E-mail:byhou@phy.nwu.edu.cn}\hspace{5mm}
      Dantao Peng$^1$ \thanks{E-mail:dtpeng@phy.nwu.edu.cn}\hspace{5mm}
      Ruihong Yue$^{1,2}$ \thanks{E-mail:yue@phy.nwu.edu.cn}\\[.3cm]
      $^1$Institute of Modern Physics Northwest University\\
       710069 Xi'An, P.R.China\\
      $^2$Institute of Theoretical Physics, Academica Science\\
      Beijing 100080, P.R.China}
\date{}
\maketitle
\begin{abstract}
The Bethe ansatz equations of the 1-D $SU(3)$ Hubbard model are
systematically derived by diagonalizing the inhomogeneous transfer matrix
of the $XXX$ model. We first derive the scattering matrix of the $SU(3)$
Hubbard model through the coordinate Bethe ansatz method. Then, with the
help quantum inverse scattering method we solve the nested transfer matrix
and give the eigenvalues, the eigenvectors and the Bethe ansatz equations.
Finally, we obtain the exactly analytic solution for the ground state.

\vspace{.5cm}


\end{abstract}

\bigskip

\setcounter{equation}{0}

\section{Introduction}

Strongly correlated electron systems have been an important research
subject in the condensed matter physics and mathematical physics. One of
the significant models is the 1-D Hubbard model for which the exact
solution was first given by Lieb and Wu\cite{Lieb}. Based on the Bethe
ansatz equations (Lieb-Wu's equations), the excited spectrum (spin
excitation and charge  excitation) was discussed in 
Refs.\cite{AO,GV,CF,FW,AAJ}. The magnetic properties of the 1-D Hubbard
model at zero temperature were investigated in Refs.\cite{MT1,HS,FK}, and
the thermodynamics of the model was studied in Refs.\cite{MT2,TK,NTA} with
the help of string hypothesis. The critical exponents of the system was
found in Refs.\cite{HV,JPA}. All physical properties can be discussed in
the framework of Bethe ansatz equations. Although there are lots of works
on Hubbard model, the integrability of  the 1-D Hubbard model was finished
until 1986 by Shastry\cite{Shastry}, Olmedilla and Wadati \cite{OW}.
Moreover, the eigenvalue of the transfer matrix related to the Hubbard
model was suggested in Ref.\cite{Shastry} and proved through different
method in Ref.\cite{Martins2,RT}. Besides, the integrability and the
exact solution of the  1-D Hubbard model with open boundary condition have
been investigated by several authors.\cite{HM,TR,SW}.

Recently, based on the Lie algebra knowledge, Maassarani and Mathieu
constructed the Hamiltonian of the $SU(n)$ XX model and shown its 
integrability\cite{MM}. Considering two coupled $SU(n)$ XX models,
Maassarani succeeded in generalizing Shastry's method to construct the
$SU(n)$ Hubbard model\cite{Ma1}. Furthermore, he found the related 
$R$-matrix which ensures the integrability of the one-dimensional $SU(n)$
Hubbard model\cite{Ma2}.(It was also proved by Martins for $n = 3, 
4$.\cite{Martins1}, and by Yue amd Sasaki\cite{yuesasaki} for general $n$
in terms of Lax-pair formalism.) However, the eigenvalue and the
eigenvectors of the $SU(n)$ Hubbard model have not been given yet.

In this paper, we first apply the coordinate Bethe ansatz method to
construct the wave function and the scattering matrix of the $SU(3)$
Hubbard model in section 2, then with the help of the Yang-Baxter relation
we list the algebra consisting the elements of the transfer matrix in
section 3, which is the key relations for finding the solution of the
nested transfer matrix. In section 4 we apply the algebraic Bethe ansatz
method to discuss the eigenvalue and the Bethe ansatz equations of the
model. Then the exact solution of the $SU(3)$ Hubbard model was given out.
Under the thermodynamical limit, the explicitly analytic form of the  
ground state energy is given in section 5. In section 6, we make some
conclusions and list some questions to be considered.

\section{The Coordinate Bethe Ansatz}

The Hamiltonian of the $SU(n)$ Hubbard model is
\begin{equation}
\label{Hamiltonian}
H = \sum_{k=1}^{L}\sum_{\alpha=1}^{n-1}(E_{\sigma, k}^{n \alpha}
    E_{\sigma, k+1}^{\alpha n} + E_{\sigma, k}^{\alpha n}
    E_{\sigma, k+1}^{n \alpha} + E_{\tau, k}^{n \alpha}
    E_{\tau, k+1}^{\alpha n} + E_{\tau, k}^{\alpha n}
    E_{\tau, k+1}^{n \alpha} + \frac{U n^2}{4}\sum_{k=1}^{L}
    C_{\sigma, k}C_{\tau, k}
\end{equation}
where $U$ is the coulumb coupling constant, and $ E_{a, k}^{\alpha \beta}
(a = \sigma, \tau)$ is a matrix with zeros everywhere except for an one at
the intersection of row $\alpha$ and column $\beta$:
\begin{equation}
(E^{\alpha \beta})_{l m} = \delta_{l}^{\alpha}\delta_{m}^{\beta},
\end{equation}
The subscripts $a$ and $k$ stand for two different $E$ operators at site 
$k(k =1,\cdots,L)$. The $n \times n$ diagonal matrix $C$ is defined by $C
= \sum_{\alpha < n}E^{\alpha \alpha} - E^{n n}$. We have also assumed the
periodic boundary condition $E^{\alpha \beta}_{k+L} = E^{\alpha \beta}_k$.

It was shown that the Hamiltonian (\ref{Hamiltonian})  has a
$(su(n-1)\oplus u(1))_{\sigma}\oplus(su(n-1)\oplus u(1))_{\tau}$ symmetry.
The generators are
\begin{equation}
\label{generator}
J_{a}^{\alpha \beta} = \sum_{k=1}^{n}E_{a, k}^{\alpha \beta}
\end{equation}
and
\begin{equation}
K_a = \sum_{k=1}^L C_{a, k},
\hspace{1cm} \alpha, \beta = 1,\cdots,n-1,
\hspace{1cm} a = \sigma,\tau.
\end{equation}
It is worthy to point out that the system enjoys the $SO(4)$ symmetry when
$n=2$\cite{YS}. But the generator are different from Eq. (\ref{generator}).
In this paper, we just limit our attention on the $SU(3)$ model.

Before proceeding the coordinate Bethe ansatz approach, we first begin by
introducing some notations of our analysis. In $SU(3)$ Hubbard
model, there are two types of particle named $\sigma$ and $\tau$, and each
type of particles can occupy one of two possible states. We denote
$|0\rangle$ the vacuum state, $|1\rangle$ and $|2\rangle$ the two particle
states respectively. Under the appropriate basis
\begin{equation}
|1\rangle = \left (
\begin{array}{c}
1\\
0\\
0\\
\end{array} \right ),~~~
|2\rangle = \left (
\begin{array}{c}
0\\
1\\
0\\
\end{array} \right),~~~
|0\rangle = \left (
\begin{array}{c}
0\\
0\\
1\\
\end{array} \right )
\end{equation}
we can prove that $E^{\alpha 3}$ and $E^{3 \alpha}$ act as a creating and
a destroying operators of $|\alpha \rangle$ state respectively. 

In the coordinate Bethe ansatz method, the eigenstates of the Hamiltonian
can be assumed as
\begin{equation}
|\psi_{N_0}\rangle = \sum_{x_1\leq x_2\leq\cdots\leq x_{N_0} = 1}^L
f_{\sigma_1 \sigma_2 \cdots \sigma_{N_0}}^{\alpha_1 \alpha_2 \cdots
\alpha_{N_0}} E_{\sigma_1 x_1}^{\alpha_1 3}E_{\sigma_2 x_2}^{\alpha_2
3}\cdots E_{\sigma_{N_0} x_{N_0}}^{\alpha_{N_0} 3}|0\rangle,
\end{equation}
where
\begin{eqnarray}
f_{\sigma_1 \sigma_2 \cdots \sigma_{N_0}}^{\alpha_1 \alpha_2 \cdots
\alpha_{N_0}} & = & \sum_{P, Q}\epsilon_P\epsilon_Q A_{\sigma_{Q_1}
\sigma_{Q_2} \cdots \sigma_{Q_{N_0}}}^{\alpha_{Q_1} \alpha_{Q_2} \cdots
\alpha_{Q_{N_0}}}(k_{P_{Q_1}}, k_{P_{Q_2}}, \cdots,
k_{P_{Q_{N_0}}})\nonumber\\
& & \times \theta(x_{Q_1}\leq x_{Q_2}\leq\cdots\leq x_{Q_{N_0}})\exp(i
\sum_{j = 1}^{N_0}k_{P_j}x_{Q_j}),
\end{eqnarray}
Here $\alpha_i = 1, 2 (i=1,2,\cdots, N_0)$ stands for the different
particle states, $x_i$ the position of the particle, $\sigma_i =
\sigma, \tau$ the type of $i$th particle. The summation $P$ and $Q$ are
taken over all permutations of $N_0$ momenta $k_j$ and $N_0$ coordinates
$x_j$ respectively.The symbols $\epsilon_P$ and $\epsilon_Q$ are the
parities of two kinds of permutations. Substituting the wave function into
the Schrodinger equation
\begin{equation}
\label{Schrodinger}
H |\psi_{N_0}\rangle = E |\psi_{N_0}\rangle,
\end{equation}
we can get
\begin{equation}
A_{\cdots, \sigma_i, \sigma_j, \cdots}^{\cdots, \alpha_i, \alpha_j,
\cdots}(\cdots, k_i, k_j, \cdots) = S_{\sigma_i \sigma_j}^{\alpha_i
\alpha_j}(\sin k_i, \sin k_j)A_{\cdots, \sigma_j, \sigma_i, 
\cdots}^{\cdots, \alpha_j, \alpha_i, \cdots}(\cdots, k_j, k_i, \cdots).
\end{equation}
with $S_{\sigma_i\sigma_i}^{\alpha_i\alpha_j}(\sin k_i, \sin k_j)$ being
the two-particle scattering matrix:
\begin{equation}
\label{scattering}
S_{\sigma_i \sigma_j}^{\alpha_i \alpha_j}(\sin k_i, \sin k_j) =
\frac{\sin k_i - \sin k_j + i \gamma P_{\sigma_i \sigma_j}^{\alpha_i
\alpha_j}}{\sin k_i - \sin k_j + i \gamma}
\end{equation}
where $\gamma = \frac{9 U}{2}$, $P_{\sigma_i \sigma_j}^{\alpha_i 
\alpha_j}$ is the direct product of two kinds of permutation operators and
 permutes the particle styles and particle states simultaneously.
The energy of the Hamiltonian on this wave function is
\begin{equation}
E = 2 \sum_{i=1}^{N_0}\cos k_i + \frac{\gamma}{2}(L-2{N_0}).
\end{equation}

For the convenience, we denote by $\zeta_0$ the amplitude $A_{\cdots,
\sigma_j, \sigma_l, \cdots}^{\cdots, \alpha_j, \alpha_l, \cdots}(\cdots,
k_j, k_l, \cdots)$. When the $j$-th particle moves across the else 
particle, it gets an $S$-matrix $S_{jl}(q_j - q_l)$. Here $q_j = \sin
k_j$. The periodic boundary condition leads the following constrains on
the amplitude $\zeta_0$
\begin{equation}
\label{eigenvalue}
S_{j j+1}S_{j j+2}\cdots S_{j N_0}S_{j 1}S_{j 2}\cdots S_{j j-1}\zeta_0 =
e^{i k_j L}\zeta_0.
\end{equation}
Eq.(\ref{eigenvalue}) is similar with the Yang's eigenvalue
problem\cite{Yang}. Its solution will give out the Bethe ansatz equation.

\section{The foundamental commutation rules}

In the above section, we have obtained the scattering matrix. It can be 
proved that they satisfy the Yang-Baxter equation\cite{Yang}
\begin{equation}
S_{j l}(q_j, q_l)S_{j k}(q_j, q_k)S_{l k}(q_l, q_k) = S_{l k}(q_l,
q_k)S_{j k}(q_j, q_k)S_{j l}(q_j, q_l).
\end{equation}
Define $\zeta = (S_{j j+1}S_{j j+2}\cdots S_{j N_0})^{-1}\zeta_0$, we can
rewrite the eigenvalue problem (\ref{eigenvalue}) as
\begin{equation}
S_{j 1}S_{j 2}\cdots S_{j N_0}\zeta = X_j \zeta = \epsilon(q_j)\zeta.
\end{equation}
We introduce an auxiliary space $\tau$ and $S_{\tau j}=S_{\tau j}(q -
q_j)$, and define the monodromy matrix
\begin{equation}
T_{\tau}(q)=S_{\tau 1}S_{\tau 2}\cdots S_{\tau N_0}
\end{equation}
Obviously, $S_{\tau j}(q_j) = P_{\tau j}$ and
\begin{equation}
tr_\tau T_\tau(q_j)=X_j.
\end{equation}
Since $S_{\tau i}$ satisfies the Yang-Baxter equation, we can easily prove
that the monodromy matrix $T_\tau(q)$ also satisfies the Yang-Baxter
relation
\begin{equation}
S_{\tau \tau^\prime}(q, q^\prime)T_{\tau j}(q, q_j)T_{\tau^\prime
j}(q^\prime, q_j) = T_{\tau^\prime j}(q^\prime, q_j)T_{\tau j}(q,
q_j)S_{\tau \tau^\prime}(q, q^\prime).
\end{equation}

From the point of view of a vertex model, we can interpret the matrix
$S_{\tau j}$ as the vertex operator, the matrix $S_{\tau \tau^\prime}$ as
the $R$-matrix. So the $R$-matrix is an $16\times 16$ matrix
\begin{equation}
R_{j l}(q_j, q_l) = \left (
\begin{array}{llllllllllllllll}
\alpha_1 & 0 & 0 & 0 & 0 & 0 & 0 & 0 & 0 & 0 & 0 & 0 & 0 & 0 & 0 & 0 \\
0 & \alpha_2 & 0 & 0 & \alpha_3 & 0 & 0 & 0 & 0 & 0 & 0 & 0 & 0 & 0 & 0 &
0 \\
0 & 0 & \alpha_2 & 0 & 0 & 0 & 0 & 0 & \alpha_3 & 0 & 0 & 0 & 0 & 0 & 0 &
0 \\
0 & 0 & 0 & \alpha_2 & 0 & 0 & 0 & 0 & 0 & 0 & 0 & 0 & \alpha_3 & 0 & 0 &
0 \\
0 & \alpha_3 & 0 & 0 & \alpha_2 & 0 & 0 & 0 & 0 & 0 & 0 & 0 & 0 & 0 & 0 &
0 \\
0 & 0 & 0 & 0 & 0 & \alpha_1 & 0 & 0 & 0 & 0 & 0 & 0 & 0 & 0 & 0 & 0 \\
0 & 0 & 0 & 0 & 0 & 0 & \alpha_2 & 0 & 0 & \alpha_3 & 0 & 0 & 0 & 0 & 0 &
0 \\
0 & 0 & 0 & 0 & 0 & 0 & 0 & \alpha_2 & 0 & 0 & 0 & 0 & 0 & \alpha_3 & 0 &
0 \\
0 & 0 & \alpha_3 & 0 & 0 & 0 & 0 & 0 & \alpha_2 & 0 & 0 & 0 & 0 & 0 & 0 &
0 \\
0 & 0 & 0 & 0 & 0 & 0 & \alpha_3 & 0 & 0 & \alpha_2 & 0 & 0 & 0 & 0 & 0 &
0 \\
0 & 0 & 0 & 0 & 0 & 0 & 0 & 0 & 0 & 0 & \alpha_1 & 0 & 0 & 0 & 0 & 0 \\
0 & 0 & 0 & 0 & 0 & 0 & 0 & 0 & 0 & 0 & 0 & \alpha_2 & 0 & 0 & \alpha_3 &
0 \\
0 & 0 & 0 & \alpha_3 & 0 & 0 & 0 & 0 & 0 & 0 & 0 & 0 & \alpha_2 & 0 & 0 &
0 \\
0 & 0 & 0 & 0 & 0 & 0 & 0 & \alpha_3 & 0 & 0 & 0 & 0 & 0 & \alpha_2 & 0 &
0 \\
0 & 0 & 0 & 0 & 0 & 0 & 0 & 0 & 0 & 0 & 0 & \alpha_3 & 0 & 0 & \alpha_2 &
0 \\
0 & 0 & 0 & 0 & 0 & 0 & 0 & 0 & 0 & 0 & 0 & 0 & 0 & 0 & 0 & \alpha_1 \\

\end{array}
\right )
\end{equation}
where
\begin{equation}
\alpha_1(q_j, q_l) = 1,\hspace{.5cm}
\alpha_2(q_j, q_l) = \frac{q_j - q_l}{q_j - q_l + i\gamma},\hspace{.5cm}
\alpha_3(q_j, q_l) = \frac{i\gamma}{q_j - q_l + i\gamma}.
\end{equation}
The vertex operator is
\begin{eqnarray}
{\cal L}_{\tau j}(q - q_j) & = & S_{\tau j}(q - q_j)\nonumber\\
& = & \alpha_2(q, q_j) + \alpha_3(q, q_j)\left\{\frac{1}{2}(1 +
\sigma_\tau^z \sigma_j^z) + \sigma_\tau^+ \sigma_j^- + \sigma_{\tau}^-
\sigma_j^+\right\} \nonumber\\
& & \otimes \left\{\frac{1}{2}(1 + \tau_\tau^z \tau_j^z) + \tau_\tau^+
\tau_j^- + \tau_\tau^- \tau_j^+\right\},
\end{eqnarray}
where $\sigma_j, \tau_j$ are two different kinds of Pauli matrixes on
$j$-th space.

In addition to the $L$-operator and the $R$-matrix the existence of a
local reference state is another important object in the quantum inverse
scattering program. In our case, we can define the local reference
state by
$$
|0\rangle_j = \left (
\begin{array}{c}
1\\
0\\
\end{array}\right )_{\sigma_j} \otimes \left (
\begin{array}{c}
1\\
0\\
\end{array}\right )_{\tau_j} = |\uparrow\rangle_{\sigma_j} \otimes
|\uparrow\rangle_{\tau_j}
$$
The action of $L$-operator on this state has the
following property:
\begin{equation}
\label{L-property}
{\cal L}_{\tau j}(q ) |0\rangle_j = \left (
\begin{array}{cccc}
|0\rangle_j & |\uparrow\downarrow\rangle_j &
|\downarrow\uparrow\rangle_j & |\downarrow\downarrow\rangle_j \\[3mm]
0 & \displaystyle\frac{q}{  q+i\gamma}|0\rangle_j & 0 & 0 \\[3mm]
0 & 0 &\displaystyle \frac{q}{  q+i\gamma}|0\rangle_j & 0 \\[3mm]
0 & 0 & 0 & \displaystyle\frac{q}{  q+i\gamma}|0\rangle_j 
\end{array}\right )
\end{equation}

The global reference state $|0\rangle$ is then defined by the tensor
product of local reference states, i.e. $|0\rangle =
\prod_{j=1}^L\otimes|0\rangle_j$. This state is an eigenstate of the
transfer matrix which is the trace of monodromy matrix. In order to
construct the other eigenstates, it is necessary to seek for an 
appropriate representation of the monodromy matrix. By this we mean a
structure which is able to distinguish creation and annihilation operators
as well as possible hidden symmetries. The property of L-operator suggests
the monodromy matrix to take the following form\cite{Martins2}
\begin{equation}
T_\tau(q) = \left (
\begin{array}{lll}
B(q) & {\bf B}(q) & F(q) \\
{\bf C}(q) & {\bf A}(q) & {\bf B^*}(q) \\
C(q) & {\bf C^*}(q) & D(q)
\end{array}\right )_{4\times 4}
\end{equation}
where ${\bf B}(q),{\bf C^*}(q)$ and ${\bf B^*}(q), {\bf C}(q)$ are two
component vectors with dimensions $1\times 2$ and $2\times 1$
respectively. The operator ${\bf A}(q)$ is a $2\times 2$ matrix and we
shall denote its elements by $A_{a b}(q)$. The remaining operators $B(q),
C(q), F(q), D(q)$ are scalars.

In the framework of the above partition the eigenvalue problem for the
transfer matrix becomes
\begin{equation}
\label{transferprob}
[B(q) + \sum_{a=1}^2 A_{a a}(q) + D(q)]|\Phi\rangle =
\Lambda(q)|\Phi\rangle.
\end{equation}
where $\Lambda(q)$ and $|\Phi\rangle$ correspond to the eigenvalue and the
eigenvector respectively. From Eq. (\ref{L-property}), we know the action
of the monodromy matrix on the reference state:
\begin{equation}
\label{monodromyaction}
\begin{array}{ll}
B(q)|0\rangle = |0\rangle, & D(q)|0\rangle =\omega(q)|0\rangle,\\[3mm]
A_{a a}(q)|0\rangle = \omega(q)|0\rangle, & a = 1, 2.\\[3mm]
{\bf C}(q)|0\rangle = 0, & {\bf C^*}(q)|0\rangle = 0,\\[3mm]
C(q)|0\rangle = 0, & A_{a b}(q)|0\rangle = 0,\quad a \neq b.
\end{array}
\end{equation}
where the function $\omega(q)$ is defined by 
\begin{equation}
\omega(q)=\prod_{j=1}^{N_0}\frac{q - q_j}{q - q_j + i\gamma}.
\end{equation}
The operator ${\bf B}(q)$ and $F(q)$ play the role of  creation operators
over the reference state $|0\rangle$.

To make further progress we have to recast the Yang-Baxter algebra in the
form of commutation relations for the creation and annihilation
operators. We shall start our discussion by the commutation rule between
the operators ${\bf B}(q)$ and ${\bf B}(p)$:
\begin{equation}
\label{comm-BvBv}
{\bf B}(q)\otimes{\bf B}(p) = [{\bf B}(p)\otimes{\bf B}(q)]\cdot\hat{r}(q,
p).
\end{equation}
where $\hat{r}(q, p)$ is an auxiliary $4\times 4$ matrix given by
\begin{equation}
\label{rdef}
\hat{r}(q, p)=\left (
\begin{array}{cccc}
1 & 0 & 0 & 0\\[3mm]
0 & \hat{a}(q, p) & \hat{b}(q, p) & 0\\[3mm]
0 & \hat{b}(q, p) & \hat{a}(q, p) & 0\\[3mm]
0 & 0 & 0 & 1
\end{array}\right )
\end{equation}
and the functions $\hat{a}(q, p)$ and $\hat{b}(q, p)$ are defined by
\begin{equation}
\label{rpdef}
\hat{a}(q, p) = \frac{\alpha_3(q, p)}{\alpha_1(q, p)},\hspace{1cm}
\hat{b}(q, p) = \frac{\alpha_2(q, p)}{\alpha_1(q, p)}
\end{equation}

We can see that this auxiliary matrix $r(q, p)$ is precisely the
rational $R$-matrix of the isotropic $6$-vertex model or the XXX spin
chain. It is very easy to see out that
\begin{equation}
\hat{a}(q, p) = 1 - \hat{b}(q, p),
\end{equation}

To solve the eigenvalue problem (24) we need several other commutation
rules, especially the commutation rules between the diagonal and
the creation
operators. The commutation relations between the diagonal and creation
operator ${\bf B}(q)$ are
\begin{equation}
\label{comm-ABv}
{\bf A}(q)\otimes{\bf B}(p) = \frac{\alpha_1(q, p)}{\alpha_2(q, p)}[{\bf
B}(p)\otimes{\bf A}(q)]\cdot\hat{r}(q, p) - \frac{\alpha_3(q, p)}
{\alpha_2(q, p)}{\bf B}(q)\otimes{\bf A}(p),
\end{equation}
\begin{equation}
\label{comm-BBv}
B(q){\bf B}(p) = \frac{\alpha_1(p, q)}{\alpha_2}{\bf B}(p)B(q) -
\frac{\alpha_3(p, q)}{\alpha_2(p, q)}{\bf B}(q)B(p),
\end{equation}
\begin{equation}
\label{comm-DBv}
D(q){\bf B}(p) = {\bf B}(p)D(q) + \frac{\alpha_3(q, p)}{\alpha_2(q,
p)}F(p){\bf C^*}(q) - \frac{\alpha_3(q, p)}{\alpha_2(q, p)}F(q){\bf
C^*}(p)
\end{equation}
We also need the following commutation rule
\begin{equation}
\label{comm-CvBv}
{\bf C^*}(q)\otimes{\bf B}(p) = \frac{\alpha_1(q, p)}{\alpha_2(q, p)}[{\bf
B}(p)\otimes{\bf C^*}(q)]\cdot\hat{r}(q, p) - \frac{\alpha_3(q,
p)}{\alpha_2(q, p)}{\bf B}(q)\otimes{\bf C^*}(p)
\end{equation}

We have set up the basic tools to construct the eigenvectors of the
eigenvalue problem (\ref{transferprob}). In the next section we will show
how this problem can be solved with the help of the commutations rules
(\ref{comm-BvBv}), (\ref{rdef}), (\ref{comm-ABv})-(\ref{comm-CvBv}).

\section{The eigenvectors and eigenvalue construction}

The purpose of this section is to solve the eigenvalue problem for the
transfer matrix. We shall begin by considering the construction of an
ansatz for the corresponding eigenvectors.

\subsection{The eigenvalue problem}

The eigenvectors of the transfer matrix are in principle built up in terms
of a linear combination of the products of  many creation operators
acting
on the reference state, which are characterized by a set of rapidities
parameterizing the creation operators. First, we define the eigenvector
for an arbitrary $N_1$-particle state:
\begin{equation}
\label{eigenvector}
|\Phi_{N_1}(\{p_l\})\rangle = {\bf \Phi}_{N_1}(p_1, p_2, \cdots,
p_{N_1})\cdot {\bf\cal F}|0\rangle
\end{equation}
where the mathematical structure of the vector ${\bf \Phi}_{N_1}(p_1, p_2,
\cdots, p_{N_1})$ will be described in terms of the creation operators. We
denote the components of vector ${\bf \cal F}$ by ${\bf\cal F}^{a_1\cdots
a_{N_1}}$ which will be determined later on, where the index $a_i$ runs
over two possible values $a_i = 1, 2$.

To construct the eigenvectors of the transfer matrix, it is sufficient to
look for combinations between the operators ${\bf B}(q)$ and $F(q)$. The
experience on constructing a few particle states suggests us the 
eigenvector of $N_1$-particle to be  
\begin{equation}
{\bf \Phi}_{N_1}(p_1, \cdots, p_{N_1}) = {\bf B}(p_1)\otimes{\bf
\Phi}_{N_1 - 1}(p_2, \cdots, p_{N_1}).
\end{equation}
Here we have formally identified ${\bf \Phi}_0$ with the unity vector. The
detailed construction for less than $3$ particles was given in Appendix B.
This vector has the following symmetry:
\begin{equation}
{\bf \Phi}_{N_1}(p_1, \cdots, p_{j-1}, p_j, \cdots, p_{N_1}) = {\bf
\Phi}_{N_1}(p_1, \cdots, p_j, p_{j-1}, \cdots, p_{N_1})\cdot\hat{r}_{j-1
j}(p_{j-1}, p_j), \end{equation}
where the subscripts in $\hat{r}_{j-1 j}(p_{j-1}, p_j)$ emphasize the
positions in the $N_1$-particle space $V_1\otimes\cdots\otimes
V_{j-1}\otimes V_j\otimes\cdots\otimes V_{N_1}$ on which this matrix acts
non-trivially. Here we have already assumed that the $(N_1 - 1)$-particle
state was already symmetrized. 

Applying the diagonal elements of monodromy matrix on this eigenstate, we
can get
\begin{eqnarray}
\label{B-action}
B(q) |\Phi_{N_1}(\{p_l\})\rangle & = &
\prod_{i=1}^{N_1}\frac{\alpha_1(p_i, q)}{\alpha_2(p_i, q)} 
|\Phi_{N_1}(\{p_l\})\rangle\nonumber\\ 
& & - \sum_{i=1}^{N_1}\frac{\alpha_3(p_i, q)}{\alpha_2(p_i, q)}
\prod_{k=1, k \neq i}^{N_1}\frac{\alpha_1(p_k, p_i)}{\alpha_2(p_k, p_i)}
|\Psi_{N_1 - 1}^{(1)}(q, p_i; \{p_l\})\rangle.\\
\label{D-action}
D(q) |\Phi_{N_1}(\{p_l\})\rangle & = & \omega(q)
|\Phi_{N_1}(\{p_l\})\rangle.\\
\label{A-action}
\sum_{a=1}^2 A_{a a}(q)|\Phi_{N_1}(\{p_l\})\rangle & = &
\omega(q)\prod_{i=1}^{N_1}\frac{\alpha_1(q, p_i)}{\alpha_2(q, p_i)}
\Lambda^{(1)}(q, \{p_l\})|\Phi_{N_1}(\{p_l\})\rangle\nonumber \\
& & - \sum_{i=1}^{N_1}\omega(p_i)\frac{\alpha_3(q, p_i)}{\alpha_2(q, p_i)}
\prod_{k=1, k \neq i}^{N_1}\frac{\alpha_1(p_i, p_k)}{\alpha_2(p_i, p_k)}
\Lambda^{(1)}(q, \{p_l\})\nonumber\\
& & \times|\Psi_{N_1 - 1}^{(1)}(q, p_i, \{p_l\})\rangle.
\end{eqnarray} 
here
\begin{equation}
\label{func-def}
|\Psi_{N_1 - 1}^{(1)}(q, p_i, \{p_l\})\rangle = {\bf B}(q)\otimes{\bf
\Phi}_{N_1 - 1}(p_1, \cdots, \check{p}_i, \cdots, 
p_{N_1})\hat{O}_i^{(1)}(p_i, \{p_l\})\cdot{\bf\cal F}|0\rangle.
\end{equation}
and
\begin{equation}
\label{Ofunc-def}
\hat{O}_i^{(1)}(p_i, \{p_k\})=\prod_{k=1}^{i-1}\hat{r}_{k, k+1}(p_k, p_i)
\end{equation}
where the symbol $\check{p}_i$ means that the rapidity $p_i$ is absent
from the set $\{p_1, \cdots, p_{N_1}\}$.

The terms proportional to the eigenvector $|\Phi_{N_1}(\{p_l\})\rangle$
are denominated the wanted terms because they contribute directly to the
eigenvalue. From Eqs. (\ref{B-action})-(\ref{Ofunc-def}), we can directly
get the eigenvalue of $N_1$-particle state
\begin{equation}
\label{tran-value}
\Lambda(q, \{p_l\}) = \prod_{i=1}^{N_1}\frac{\alpha_1(p_i,
q)}{\alpha_2(p_i, q)} + \omega(q) + 
\omega(q)\prod_{i=1}^{N_1}\frac{\alpha_1(q,  p_i)}{\alpha_2(q,
p_i)}\Lambda^{(1)}(q, \{p_l\}).
\end{equation}
The remaining ones are called unwanted terms and can be eliminated
by imposing further restrictions on the rapidities $p_i$. These
restrictions, known as the Bethe Ansatz equations, are
\begin{equation}
\label{tran-BA}
\frac{1}{\omega(p_i)} = \prod_{k=1, k \neq i}^{N_1}\frac{\alpha_2(p_k,
p_i)}{\alpha_2(p_i, p_k)}\Lambda^{(1)}(p_i, \{p_l\}). \hspace{1cm}i = 1,
\cdots, n.
\end{equation}

In fact, the undeterminated eigenvalue $\Lambda^{(1)}(q, \{p_l\})$ must
satisfy the following auxiliary problem
\begin{equation}
\label{auxpro}
T^{(1)}(q, \{p_l\})^{b_1\cdots b_{N_1}}_{a_1\cdots a_{N_1}}{\cal
F}^{b_1\cdots b_{N_1}} = \Lambda^{(1)}(q, \{p_l\}){\cal F}^{a_1\cdots
a_{N_1}},
\end{equation}
where the inhomogeneous transfer matrix $T^{(1)}(q, \{p_l\})$ is
\begin{equation}
T^{(1)}(q, \{p_l\})^{b_1\cdots b_{N_1}}_{a_1\cdots a_{N_1}} = \hat{r}^{c_1
b_1}_{a_1 d_1}(q, p_1)\hat{r}^{d_1 b_2}_{a_2 d_2}(q, 
p_2)\cdots\hat{r}^{d_{N_1 - 1} b_{N_1}}_{a_{N_1} c_1}(q, p_{N_1}).
\end{equation}

This result is the direct extensions of that obtained from the 
two-particle state. The Bethe ansatz equations and the eigenvalue still
depend on this additional auxiliary eigenvalue problem.

\subsection{The eigenvalues and the nested Bethe ansatz}

The purpose of this subsection will be the diagonalization of the
auxiliary
transfer matrix $T^{(1)}(q, \{p_l\})$. First, we write the transfer matrix
$T^{(1)}(q, \{p_l\})$ as the trace of
the following monodromy matrix:
\begin{equation}
{\cal T}^{(1)}(q, \{p_l\}) = {\cal L}^{(1)}_{{\cal A}^{(1)}N_1}(q,
p_{N_1}){\cal L}^{(1)}_{{\cal A}^{(1)}N_1-1}(q, p_{N_1-1})\cdots{\cal
L}^{(1)}_{{\cal A}^{(1)}1}(q, p_1).
\end{equation}
where ${\cal A}^{(1)}$ is the auxiliary space. The $L$-operator ${\cal
L}^{(1)}_{{\cal A}^{(1)}i}(q, p_i)$ is related to the auxiliary matrix
$\hat{r}(q, p_i)$ by a permutation operator
\begin{equation}
{\cal L}^{(1)}_{{\cal A}^{(1)}i}(q, p_i) = \left (
\begin{array}{cccc}
1 & 0 & 0 & 0\\[3mm]
0 & \hat{b}(q, p_i) & \hat{a}(q, p_i) & 0\\[3mm]
0 & \hat{a}(q, p_i) & \hat{b}(q, p_i) & 0\\[3mm]
0 & 0 & 0 & 1
\end{array}\right )
\end{equation}

Comparing with $6$-vertex model, one can find that this problem is
equivalent to the $6$-vertex model. So we can take the
results of Ref.\cite{Martins2} directly.
The eigenvalue of the auxiliary problem \ref{auxpro} is:
\begin{equation}
\label{Aux-value}
\Lambda^{(1)}(q, \{p_l\} \{s_k\}) = \prod_{k=1}^{N_2} 
\frac{1}{\hat{b}(s_k, q)} + \prod_{l=1}^{N_1}\hat{b}(q,
p_l)\prod_{k=1}^{N_2}\frac{1}{\hat{b}(q, s_k)}.
\end{equation}
and the parameters $\{s_k\}$ satisfy the following restrictions:
\begin{equation}
\label{Aux-BA}
\prod_{l=1}^{N_1}\hat{b}(s_k, p_l) = \prod_{i=1}^{N_2}\frac{\hat{b}(s_k,
s_i)}{\hat{b}(s_i, s_k)}, \hspace{1cm} k = 1, \cdots, {\cal M}. 
\end{equation}

Now we use the auxiliary eigenvalue expression to rewrite our previous
results for the eigenvalues and the Bethe ansatz equations. Substituting
the expression (\ref{Aux-value}) in Eqs.(\ref{tran-value},\ref{tran-BA})
and using the relation (\ref{rpdef}) we obtain that the eigenvalue is
\begin{eqnarray}
\Lambda(q, \{p_l\}, \{s_k\}) & = & \prod_{l=1}^{N_1}\frac{(p_l - q +
i\gamma)}{(p_l - q)} + \prod_{j=1}^{N_0}\frac{(q - q_j)}{(q - q_j +
i\gamma)} + \prod_{j=1}^{N_0}\frac{(q - q_j)}{(q - p_j + 
i\gamma)}\nonumber\\
& & \times\left\{\prod_{l=1}^{N_1}\frac{(q - p_l + i\gamma)}{(q -
p_l)}\prod_{k=1}^{N_2}\frac{(s_k - q + i\gamma)}{(s_k- q)} +
\prod_{k=1}^{N_2}\frac{(q - s_k + i\gamma)}{(q - s_k)}\right\}
\end{eqnarray}
In terms of the above eigenvalue, the periodic boundary condition
(\ref{eigenvalue}) implies
\begin{equation}
\label{periodic-BA}
e^{i k_j L} = \Lambda(q_j, \{p_l\}, \{s_k\}) =
\prod_{l=1}^{N_1}\frac{(\sin k_j - \tilde{p_l} - i\frac{\gamma}{2})}{(\sin
k_j - \tilde{p_l} + i \frac{\gamma}{2})}
\end{equation}
The second set of Bethe ansatz equations Eq. (\ref{tran-BA}) for the 
rapidities $p_l$ changes into
\begin{equation}
\label{Second-BA}
\prod_{i=1, i \neq l}^{N_1}\frac{(\tilde{p_l} - \tilde{p_i} + i 
\gamma)}{(\tilde{p_l} - \tilde{p_i} - i \gamma)} =
\prod_{j=1}^{N_0}\frac{(\tilde{p_l} - \sin k_j + i
\frac{\gamma}{2})}{(\tilde{p_l} - \sin k_j - i
\frac{\gamma)}{2})}\prod_{k=1}^{N_2}\frac{(\tilde{p_l} -
\tilde{s_k} + i \frac{\gamma}{2})}{(\tilde{p_l} - \tilde{s_k} - i
\frac{\gamma}{2})}, \hspace{.5cm}(l = 1, 2, \cdots, M).
\end{equation}
and the third set of Bethe ansatz equations Eq. (\ref{Aux-BA}) for the
rapidities $s_k$ is
\begin{equation}
\label{Third-BA}
\prod_{l=1}^{N_1}\frac{(\tilde{s_k} - \tilde{p_l} - i
\frac{\gamma}{2})}{(\tilde{s_k} - \tilde{p_l} + i \frac{\gamma)}{2})} =
\prod_{i=1, i \neq k}^{N_2}\frac{(\tilde{s_k} - \tilde{s_i} - i
\gamma)}{(\tilde{s_k} - \tilde{s_i} + i \gamma)}, \hspace{.5cm}(k = 1, 2,
\cdots, N_2).
\end{equation}
Here we have used the shifted parameters $\tilde{p_l} = p_l - 
i\frac{\gamma}{2}$ adn $ \tilde{s_k} =  s_k - i \gamma$ to bring our
equations to more symmetric forms.

\section{The ground state}

In the proceeding section, we have obtained the final Bethe ansatz
equations of the $SU(3)$ Hubbard model. In this section, we will use them
to analyse the ground state property of the model.

By taking the logarithm of the Bethe ansatz equations 
(\ref{periodic-BA})-(\ref{Third-BA}), we can get the following three
sets of equations (In this section, we assume $U > 0$):
\begin{equation}
k_j L = 2\pi I_j + 2\sum_{l=1}^{N_1}\tan^{-1}\left(\frac{\sin k_j
-  \tilde{p_l}}{\frac{\gamma}{2}}\right),
\end{equation}
\begin{equation}
2\pi J_l + 2\sum_{i=1, i\neq l}^{N_1}\tan^{-1}\left(\frac{\tilde{p_l} -
\tilde{p_i}}{\gamma}\right) =   
2\sum_{j=1}^{N_0}\tan^{-1}\left(\frac{\tilde{p_l} -  
\sin{k_j}}{\frac{\gamma}{2}}\right) + 
2\sum_{k=1}^{N_2}\tan^{-1}\left(\frac{\tilde{p_l} -
\tilde{s_k}}{\frac{\gamma}{2}}\right),
\end{equation}
\begin{equation}
2\sum_{l=1}^{N_1}\tan^{-1}\left(\frac{\tilde{s_k} - 
\tilde{p_l}}{\frac{\gamma}{2}}\right) = 2\sum_{i=1, i \neq
k}^{N_2}\tan^{-1}\left(\frac{\tilde{s_k} - \tilde{s_i}}{\gamma}\right) +
2\pi P_k, 
\end{equation}
where $I_j, J_l, P_k$ are integers or half-odd integers. Under the
thermodynamic limits the Bethe ansatz equations for the ground state
change into
\begin{eqnarray}
\label{rho}
2\pi\rho(k) & = & 1 - \cos k 
\int_{-B}^B\frac{4\gamma\sigma(\Lambda)}{\gamma^2 + 4(\sin k - 
\Lambda)^2}d \Lambda,\\
\nonumber\\
\label{sigma}
2\pi\sigma(\Lambda) & + &
\int_{-B}^B\frac{2 \gamma \sigma(\Lambda^\prime)}{\gamma^2 + (\Lambda -
\Lambda^\prime)^2}d \Lambda^\prime = \nonumber\\
\nonumber\\
& & \int_{-Q}^Q\frac{4\gamma\rho(k)}{\gamma^2 + 4(\Lambda - \sin k)^2}d k
+ \int_{-E}^E\frac{4\gamma\tau(\Lambda^\prime)}{\gamma^2 + 4(\Lambda -
\Lambda^\prime)^2}d\Lambda^\prime,\\
\nonumber\\
\label{tau}
2\pi\tau(\Lambda) & + &
\int_{-E}^E\frac{2\gamma\tau(\Lambda^\prime)}{\gamma^2
+ (\Lambda - \Lambda^\prime)^2}d \Lambda^\prime = 
\int_{-B}^B\frac{4\gamma\sigma(\Lambda^\prime)}{\gamma^2 + 4(\Lambda -
\Lambda^\prime)^2}d \Lambda^\prime,
\end{eqnarray}
where $Q$,$B$ and $E$ are determined by the conditions
\begin{eqnarray}
\label{Q}
\int_{-Q}^Q\rho(k)d k = \frac{N_0}{L},\\
\nonumber\\
\label{B}
\int_{-B}^B\sigma(\Lambda)d \Lambda = \frac{N_1}{L},\\
\nonumber\\
\label{E}
\int_{-E}^E\tau(\Lambda^\prime)d \Lambda^\prime = \frac{N_2}{L}. 
\end{eqnarray}
The function $\rho(k), \sigma(\Lambda), \tau(\Lambda^\prime)$ are the 
distribution functions of real parameters $k_j, \tilde{p_l}$ and  
$\tilde{s_k}$ respectively.

Eqs. (\ref{rho})-(\ref{E}) have a unique solution which is positive for
all allowed
$Q$, $B$ and $E$. $\frac{N_i}{N}, (i=0, 1, 2)$ is a monotonically
function of $Q$, $B$ and $E$ respectively. Thus the ground state is
characterized by $Q=\pi, B=E=\infty$.

After taking Fourier transforms of the above equations we can obtain the
result of the distribution functions
\begin{eqnarray}
\rho(k) & = & \frac{1}{2\pi} - \frac{\cos{k}}{2\pi}\int_{-\infty}^{\infty}
\frac{1 + e^{-\gamma|\omega|}}{1 + 2\cosh(\gamma|\omega|)}
e^{-i\omega\sin{k}}J_0(\omega)d\omega,\\
\nonumber\\
\sigma(\Lambda) & = & 
\frac{1}{2\pi}\int_{-\infty}^{\infty}\frac{2\cosh(\frac{\omega}{2}
|\omega|)}{4\cosh^2(\frac{\gamma}{2}|\omega|) -
1}J_0(\omega)e^{-i\omega\Lambda}d\omega,\\
\nonumber\\
\tau(\Lambda^\prime) & = & \frac{1}{2\pi}\int_{-\infty}^{\infty}\frac{1}{4
\cosh^2(\frac{\gamma}{2}|\omega|) -
1}J_0(\omega)e^{-i\omega\Lambda^\prime}d\omega.
\end{eqnarray}
From Eqs.(61)-(66) we have $N_0 = L$, this means that all lattice site is
filled by one particles, $N_1 = \frac{2}{3}N_0$, and $N_2 = 
\frac{1}{3}N_0$.

The ground state energy then is given by
\begin{eqnarray}
E & = & 2 L \int_{-\infty}^{\infty}\rho(k)\cos{k}d k -
\frac{\gamma}{2}L \nonumber\\
\nonumber\\
& = & - 2 L \int_{-\infty}^{\infty}\frac{1 + 
e^{-\gamma|\omega|}}{\omega[1  +
2\cosh(\gamma|\omega|)]}J_0(\omega)J_1(\omega)d\omega - \frac{\gamma}{2}L.
\end{eqnarray}

\section{Conclusions}
The main purpose of this paper was to investigate the eigenvalue of
the $SU(3)$ Hubbard model. We have succeeded in finding the eigenvalue of
the Hamiltonian (\ref{Hamiltonian}) and obtained three sets of Bethe
ansatz equations. Based on the Bethe ansatz equations, we have found the
explicit expression of the energy and the distribution functions of the
rapidities corresponding to the ground state for positive $U$.

One important question is to study the excitation spectrum and the low
tempareture thermodynamics of the model for both positive and negative
$U$. The Bethe ansatz equations given in present paper will play an key
role. As we know, the negative $U$ has a distinguished properties from
positive $U$ case. The solution structure of Bethe ansatz equations
corresponding to the ground state and excited spectrum for attractive case
$(U<0)$ are different from those for repulsive case. It is worthy to be
studied in future.

In Ref.\cite{Ma2}, the $L$-operator and $R$-matrix were given, therefore,
one can define the transfer matrix as done in usual Hubbard model (
$SU(2)$ case). But how
to find the eigenvalue of the transfer matrix is unknown. It may be solved
by using the method proposed in Ref.\cite{Martins2}. We will consider this
problem late. Another natural generalization the present paper is to
find the eigenvalue of the Hamiltonian (\ref{Hamiltonian}) for general
$n$, it will be considered elsewhere.  

%%%%%%%%%%%%%%%%%%%%%% Appendix %%%%%%%%%%%%%%%%%%%%%%%%%%%%%%%%%%%%
\setcounter{equation}{0}
\renewcommand{\theequation}{A.\arabic{equation}}

\section{Appendix A:}

\noindent First the one-particle eigenstate can be assumed to be
\begin{equation}
|\psi_1\rangle = \sum_{x=1}^L f_{\sigma_1}^{\alpha_1}(x)E_{\sigma_1
x}^{\alpha_1 3}|0\rangle.
\end{equation}
Applying the Hamiltonian (\ref{Hamiltonian}) into this ansatz, we can
obtain
\begin{equation}
E f_{\sigma_1}^{\alpha_1}(x) = f_{\sigma_1}^{\alpha_1}(x + 1) +
f_{\sigma_1}^{\alpha_1}(x - 1) + \frac{9
U}{4}(L - 2)f_{\sigma_1}^{\alpha_1}(x), \hspace{5mm} 1 < x < L.
\end{equation}
The solution of above equation is
\begin{equation}
f_{\sigma_1}^{\alpha_1}(x) = A_{\sigma_1 +}^{\alpha_1}e^{i k x} -
A_{\sigma_1 -}^{\alpha_1}e^{- i k x},
\end{equation}
with the eigenenergy
\begin{equation}
E = 2 \cos{k} + \frac{9 U}{4}(L - 2),
\end{equation}
The periodic boundary condition governs the momentum to be
\begin{equation}
k_j = \frac{2 j \pi}{L}, \hspace{5mm} (j = 1, \cdots, L-1).
\end{equation}
 
For the two-particle state, the eigenstate is assumed as
\begin{equation}
|\psi_2\rangle = \sum_{x_1, x_2 = 1}^{L}f_{\sigma_1 \sigma_2}^{\alpha_1
\alpha_2}(x_1, x_2)E_{\sigma_1 x_1}^{\alpha_1 3}E_{\sigma_2 x_2}^{\alpha_2
3}|0\rangle.
\end{equation}
From the Schrodinger equation (\ref{Schrodinger}) we have
\begin{eqnarray}
\label{coordinate}
E f_{\sigma_1 \sigma_2}^{\alpha_1 \alpha_2}(x_1, x_2) & = & f_{\sigma_1
\sigma_2}^{\alpha_1 \alpha_2}(x_1 + 1, x_2) + f_{\sigma_1
\sigma_2}^{\alpha_1 \alpha_2}(x_1 - 1, x_2) +\frac{9 U}{4}(L
-4)f_{\sigma_1 \sigma_2}^{\alpha_1 \alpha_2}(x_1, x_2)\nonumber\\
& & + f_{\sigma_1 \sigma_2}^{\alpha_1 \alpha_2}(x_1, x_2 + 1) +
f_{\sigma_1 \sigma_2}^{\alpha_1 \alpha_2}(x_1, x_2 - 1), \hspace{5mm}
x_1 \neq x_2, \nonumber\\
E f_{\sigma_1 \sigma_2}^{\alpha_1 \alpha_2}(x_1, x_1) & = & f_{\sigma_1
\sigma_2}^{\alpha_1 \alpha_2}(x_1 + 1, x_1) + f_{\sigma_1
\sigma_2}^{\alpha_1 \alpha_2}(x_1 - 1, x_1) + \frac{9 U}{4}f_{\sigma_1
\sigma_2}^{\alpha_1 \alpha_2}(x_1, x_1)\nonumber\\
& & + f_{\sigma_1 \sigma_2}^{\alpha_1 \alpha_2}(x_1, x_1 + 1) +
f_{\sigma_1 \sigma_2}^{\alpha_1 \alpha_2}(x_1, x_1 - 1), \hspace{5mm}
x_1 = x_2.
\end{eqnarray}
The fermionic property of particle requires the antisymmetry of the wave
function. Therefore, we can assume the wave function $f_{\sigma_1
\sigma_2}^{\alpha_1 \alpha_2}(x_1, x_2)$ to be
\begin{equation}
f_{\sigma_1 \sigma_2}^{\alpha_1 \alpha_2} = \left \{
\begin{array}{c}
A_{\sigma_1 \sigma_2}^{\alpha_1 \alpha_2}(k_1, k_2)e^{i k_1 x_1 + i k_2
x_2} - A_{\sigma_1 \sigma_2}^{\alpha_1 \alpha_2}(k_2, k_1)e^{i k_2 x_1 + i
k_1 x_2}, \hspace{5mm} x_1 < x_2\\
- A_{\sigma_2 \sigma_1}^{\alpha_2 \alpha_1}(k_1, k_2)e^{i k_1 x_2 + i k_2
x_1} + A_{\sigma_2 \sigma_1}^{\alpha_2 \alpha_1}(k_2, k_1)e^{i k_2 x_2 + i
k_1 x_1}, \hspace{5mm} x_1 > x_2
\end{array}\right .
\end{equation}
Substituting the above ansatz into the Eq. (\ref{coordinate}), we can
obtain the eigenvalue of the Hamiltonian (\ref{Hamiltonian})
\begin{equation}
E = 2(\cos{k_1} + \cos{k_2}) + \frac{9 U}{4}(L - 4).
\end{equation}

Imposing the continuous condition of the wave function
\begin{equation}
\lim_{x_2 \rightarrow x_1^+}f_{\sigma_1 \sigma_2}^{\alpha_1 \alpha_2}(x_1,
x_2) = \lim_{x_1 \rightarrow x_1^-}f_{\sigma_1 \sigma_2}^{\alpha_1
\alpha_2}(x_1, x_2),
\end{equation}
we have
\begin{equation}
A_{\sigma_1 \sigma_2}^{\alpha_1 \alpha_2}(k_1, k_2) = S_{\sigma_1
\sigma_2}^{\alpha_1 \alpha_2}(\sin{k_1}, \sin{k_2})A_{\sigma_2
\sigma_1}^{\alpha_2 \alpha_1}(k_2, k_1),
\end{equation}
where
\begin{equation}
S_{\sigma_1 \sigma_2}^{\alpha_1 \alpha_2}(\sin{k_1}, \sin{k_2}) =
\frac{\sin{k_1} - \sin{k_2} + i \gamma P_{\sigma_1 \sigma_2}^{\alpha_1
\alpha_2}}{\sin{k_1} - \sin{k_2} + i \gamma}
\end{equation}
is the scattering matrix of two particles with labels $(\sigma_1,
\alpha_1)$ and $(\sigma_2, \alpha_2)$, and $\gamma = \frac{9 U}{2}$,
$P_{\sigma_1 \sigma_2}^{\alpha_1 \alpha_2}$ is the direct product of
two kinds of permutation operators.

\setcounter{equation}{0}
\renewcommand{\theequation}{B.\arabic{equation}}

\section{Appendix B:}

Following the procedure of Ref. \cite{Martins2}, the one-particle 
eigenstate is
\begin{equation}
|\Phi_1(p_1)\rangle = {\bf B}(p_1)\cdot {\cal F}|0\rangle = B_a(p_1){\cal
F}^a|0\rangle.
\end{equation}
where the repeated index means sum.

With the help of commutations (\ref{comm-ABv})-(\ref{comm-DBv}) and the
properties (\ref{comm-BvBv}), (\ref{rdef}), we
find that the one-particle state satisfies the following relations
\begin{eqnarray}
B(q)|\Phi_1(p_1)\rangle & = &
\frac{\alpha_1(p_1, q)}{\alpha_2(p_1, q)}|\Phi_1(p_1)\rangle -
\frac{\alpha_3(p_1, q)}{\alpha_2(p_1, q)}{\bf B}(q)\cdot{\bf\cal 
F}|0\rangle,\\
D(q)|\Phi_1(p_1)\rangle & = & \omega(q)|\Phi_1(p_1)\rangle,\\
\sum_{a=1}^2 A_{a a}(q)|\Phi_1(p_1)\rangle & = &
\omega(q)\frac{\alpha_1(q, p_1)}{\alpha_2(q, p_1)}\hat{r}_{c_1 a_1}^{a_1
b_1}(q, p_1)B_{c_1}(p_1){\bf\cal F}^{b_1}|0\rangle\nonumber\\
& & - \omega(p_1)\frac{\alpha_3(q, p_1)}{\alpha_2(q, p_1)}{\bf
B}(q)\cdot{\bf\cal F}|0\rangle.
\end{eqnarray}
The terms proportional to the eigenvector $|\Phi_1(p_1)\rangle$ denominate
the eigenvalue
\begin{equation}
\Lambda(q, p_1) = \frac{\alpha_1(p_1, q)}{\alpha_2(p_1, q)} + \omega(q) +
\omega(q)\frac{\alpha_1(q, p_1)}{\alpha_2(q, p_1)}\Lambda^{(1)}(q, p_1).
\end{equation}
where $\Lambda^{(1)}(q, p_1)$ is defined by the one-particle auxiliary
eigenvalue problem
\begin{equation}
T^{(1)}(q, p_1)^{a_1 b_1}_{c_1 a_1}{\cal F}^{b_1} = \hat{r}^{a_1 b_1}_{c_1
a_1}(q, p_1){\cal F}^{b_1} = \Lambda^{(1)}(q, p_1){\cal F}^{c_1}
\end{equation}
The remaining ones will be eliminated by imposing the Bethe Ansatz
equation
\begin{equation}
\omega(p_1) = 1.
\end{equation}

The two-particle state should be a composition of two single hole
excitations and a local hole pair. The former is made by tensoring two
creating operators ${\bf B}(q)$ while the latter represented by
$F(q)$. Thus, we construct the two-particle vector as:
\begin{equation}
{\bf \Phi}_2(p_1, p_2) = {\bf B}(p_1)\otimes{\bf B}(p_2) + {\bf \xi}F(p_1)
B(p_2)\hat{g}^{(0)}_0(p_1, p_2),
\end{equation}
where $\hat{g}^{(0)}_0(p_1, p_2)$ is an arbitrary function to be 
determined. The vector $\bf{\xi}$ plays the role of an "exclusion"
principle, forbidding two same particles at the same site.

Putting the diagonal elements of transfer matrix on this state, we
find that the function $\hat{g}^{(0)}_0(p_1, p_2)$ must be zero such that
the two-particle state is the eigenstate of the transfer matrix. In other
words, using the definition (\ref{eigenvector}) we have
\begin{equation}
|\Phi_2(p_1, p_2)\rangle = B_i(p_1)B_j(p_2){\cal F}^{j i}|0\rangle.
\end{equation}
The diagonal operators acting on this state gives the following
relations:
\begin{eqnarray}
\label{2relation-B}
B(q)|\Phi_2(p_1, p_2)\rangle & = & \prod_{i=1}^2\frac{\alpha_1(p_i,
q)}{\alpha_2(p_i, q)}|\Phi_2(p_1, p_2)\rangle\nonumber\\
& & - \sum_{i=1}^2\frac{\alpha_3(p_i, q)}{\alpha_2(p_i, q)}\prod_{k=1, k
\neq i}^2\frac{\alpha_1(p_k, p_i)}{\alpha_2(p_k, p_i)}|\Psi^{(1)}_1(q,
p_i; \{p_l\})\rangle.\\
\label{2relation-D}
D(q)|\Phi_2(p_1, p_2)\rangle & = & \omega(q)|\Phi_2(p_1, p_2)\rangle,\\
\label{2relation-A}
\sum_{a=1}^2 A_{a a}(q)|\Phi_2(p_1, p_2)\rangle & = &
\omega(q)\prod_{i=1}^2\frac{\alpha_1(q, p_i)}{\alpha_2(q,
p_i)}\Lambda^{(1)}(q, \{p_l\})|\Phi_2(p_1, p_2)\rangle\nonumber\\
& & - \sum_{i=1}^2\omega(p_i)\frac{\alpha_3(q, p_i)}{\alpha_2(q,
p_i)}\prod_{k=1, k \neq i}^2\frac{\alpha_1(p_i, p_k)}{\alpha_2(p_i,
p_k)}\Lambda^{(1)}(p_i, \{p_l\})\nonumber\\
& & \times |\Psi^{(1)}_1(q, p_i, \{p_l\})\rangle,
\end{eqnarray}
where we have used the two-particle auxiliary relation
\begin{equation}
T^{(1)}(q, \{p_l\})^{a_1 a_2}_{b_1 b_2}{\cal F}^{a_2 a_1} = \hat{r}^{c_1
a_1}_{b_1 d_1}(q, p_1)\hat{r}^{d_1 a_2}_{b_2 c_1}(q, p_2){\cal F}^{a_2
a_1} = \Lambda^{(1)}(q, \{p_l\}){\cal F}^{b_2 b_1},
\end{equation}
and
\begin{equation}
|\Psi^{(1)}_1(q, p_i, \{p_l\})\rangle = {\bf
B}(q)\otimes{\bf \Phi}_1(p_k)\hat{O}^{(1)}_i(p_i, \{p_k\})\cdot{\bf\cal
F}|0\rangle, \hspace{5mm} k \neq i,
\end{equation}
The vanishing of unwanted terms gives us the Bethe ansatz equation
\begin{equation}
\frac{1}{\omega(p_i)} = \prod_{k=1, k \neq i}^2\frac{\alpha_2(p_k,
p_i)}{\alpha_2(p_i, p_k)}\Lambda^{(1)}(p_i, \{p_l\}), \hspace{5mm}
i=1, 2
\end{equation}
Finally, From Eqs. (\ref{2relation-B})-(\ref{2relation-A}) we can
directly read the eigenvalue of two-particle state
\begin{equation}
\Lambda(q, \{p_l\}) = \prod_{i=1}^2\frac{\alpha_1(p_i,
q)}{\alpha_2(p_i, q)} + \omega(q) +
\omega(q)\prod_{i=1}^2\frac{\alpha_1(q, p_i)}{\alpha_2(q,
p_i)}\Lambda^{(1)}(p_i, \{p_l\}).
\end{equation}

Before constructing the general multi-particle state, let us first analyse
the symmetry of the two-particle state. From the above analysis, it is
easy to prove the two-particle vector satisfying the following exchange
property
\begin{equation}
{\bf \Phi}_2(p_1, p_2) = {\bf \Phi}_2(p_2, p_1)\cdot\hat{r}(p_1, p_2).
\end{equation}
In principle, such symmetrization mechanism can be implemented to any
multi-particle state. This indeed help us to handle the problem of
constructing a general multi-particle state ansatz. We construct the
three-particle state as
\begin{equation}
{\bf \Phi}_3(p_1, p_2, p_3) = {\bf B}(p_1)\otimes{\bf B}(p_2)\otimes{\bf
B}(p_3),
\end{equation}
satisfying
\begin{equation}
{\bf \Phi}_3(p_1, p_2, p_3) = {\bf \Phi}_3(p_1, p_3, p_2)\cdot\hat{r}_{2,
3}(p_2, p_3).
\end{equation}

We also can rewrite the three-particle state in terms of the following
recurrence relation:
\begin{equation}
{\bf \Phi}_3(p_1, p_2, p_3) = {\bf B}(p_1)\otimes{\bf \Phi}_2(p_2, p_3).
\end{equation}
This expression is rather illuminating , because it suggests that a 
general $N_1$-particle state can be expressed in terms of the $N_1 -
1$-particle state via a recurrence relation. This recursive way
not only help us to better simplify the wanted terms but also makes it
possible to gather the unwanted terms in rather closed forms. The action
of diagonal operators on the three-particle state is
\begin{eqnarray}
B(q)|\Phi_3(p_1, p_2, p_3)\rangle & = & \prod_{i=1}^3\frac{\alpha_1(p_i,
q)}{\alpha_2(p_i, q)}|\Phi_3(p_1, p_2, p_3)\rangle\nonumber\\
& & - \sum_{i=1}^3\frac{\alpha_3(p_i, q)}{\alpha_2(p_i, q)}\prod_{k=1, k
\neq i}^3\frac{\alpha_1(p_k, p_i)}{\alpha_2(p_k, p_i)}\nonumber\\
& & \times|\Psi^{(1)}_2(q, p_i, \{p_l\})\rangle,\\
D(q)|\Phi_3(p_1, p_2, p_3)\rangle & = & \omega(q)|\Phi_3(p_1, p_2,
p_3)\rangle,\\
\sum_{a=1}^2 A_{a a}(q)|\Phi_3(p_1, p_2, p_3)\rangle & = &
\omega(q)\prod_{i=1}^3\frac{\alpha_1(q, p_i)}{\alpha_2(q,
p_i)}\Lambda^{(1)}(q, \{p_l\})|\Phi_3(p_1, p_2, p_3)\rangle\nonumber\\
& & - \sum_{i=1}^3\omega(p_i)\frac{\alpha_3(q, p_i)}{\alpha_2(q,
p_i)}\prod_{k=1, k \neq i}^3\frac{\alpha_1(p_i, p_k)}{\alpha_2(p_i,
p_k)}\Lambda^{(1)}(p_i, \{p_l\})\nonumber\\
& & \times|\Psi^{(1)}_2(q, p_i, \{p_l\})\rangle.
\end{eqnarray}

In general, the knowledge of the $2$-particle and the $3$-particle results
suggests the behaviour of the $N_1$-particle state. By using the
mathematical induction we are able to determine the general structure for
the multi-particle states to be Eq. (\ref{eigenvector})

\end{document}